\documentclass[aps,pra,twocolumn,showpacs,floatfix,superscriptaddress]{revtex4}
\pdfoutput=1
\usepackage{graphicx}
\usepackage{amsfonts}
\usepackage{amsthm}
\usepackage{amsmath}
\usepackage{amssymb}
\usepackage{color}
\newcommand{\ket}[1]{\vert #1 \rangle}
\newcommand{\bra}[1]{\langle #1 \vert}

\newcommand{\ketbra}[1]{\vert #1 \rangle \langle #1 \vert}
\newcommand{\operator}[1]{\mathrm{#1}}

\theoremstyle{plain}
\newtheorem{theorem}{THEOREM}
\newtheorem{corollary}{COROLLARY}
\begin{document}

\title{The Lackadaisical Quantum Walker is NOT Lazy at all}

\author{Kun Wang}
\affiliation{State Key Laboratory for Novel Software Technology,
                Department of Computer Science and Technology,
                Nanjing University, Jiangsu 210093, China}
\author{Nan Wu}
\email[Correspondence to:~]{nwu@nju.edu.cn}
\affiliation{State Key Laboratory for Novel Software Technology,
                Department of Computer Science and Technology,
                Nanjing University, Jiangsu 210093, China}
\author{Ping Xu}
\affiliation{National Laboratory of Solid State Microstructures,
            School of Physics, Nanjing University, Jiangsu 210093, China}
\author{Fangmin Song}
\affiliation{State Key Laboratory for Novel Software Technology,
                Department of Computer Science and Technology,
                Nanjing University, Jiangsu 210093, China}
\date{\today}
\begin{abstract}
In this paper, we study the properties of lackadaisical quantum walks on a line.
This model is first proposed in~\cite{wong2015grover} as a quantum analogue of lazy random walks
where each vertex is attached $\tau$ self-loops.
We derive an analytic expression for the localization probability of 
the walker at the origin after infinite steps,
and obtain the peak velocities of the walker.
We also calculate rigorously the wave function of the walker starting from the origin
and obtain a long time approximation for the entire probability density function.
As an application of the density function,
we prove that lackadaisical quantum walks spread ballistically for arbitrary $\tau$,
and give an analytic solution for the variance of the walker's probability distribution.
\end{abstract}

\maketitle
\section{Introduction}\label{sec:introduction}

Since the seminal works by~\cite{aharonov1993quantum, meyer1996quantum, farhi1998quantum},
quantum walks have been the subject of research in two decades.
They were originally proposed as a quantum generalization of random walks~\cite{spitzer2013principles}.
Asymptotic properties such as mixing time, mixing rate and hitting time
of quantum walks on a line and on general graphs have been studied extensively
~\cite{ambainis2001one, aharonov2001quantum,
moore2002quantum, childs2004spatial, krovi2006hitting}.
Applications of quantum walks in quantum information processing have also been investigated.
Especially, quantum walks can solve the element distinctness problem
~\cite{aaronson2004quantum, ambainis2007quantum} and
perform the quantum searching~\cite{szegedy2004quantum}.
In some applications, quantum walks based algorithms can even gain
exponential speedup over all possible classical algorithms~\cite{childs2003exponential}.
The discovery of their capability for universal quantum
computations~\cite{childs2009universal, lovett2010universal}
indicates that understanding quantum walks is helpful for
better understanding quantum computing itself.
For a more comprehensive review, we refer the readers to
~\cite{kempe2003quantum, venegas2012quantum} and the references within.

Lackadaisical quantum walks (LQWs), first considered by Wong et al.~\cite{wong2015grover},
are quantum analogous of lazy random walks.
This model also generalizes three-state quantum walks on a line
~\cite{inui2005one, falkner2014weak, vstefavnak2014limit, wang2016grover},
which only have one self-loop at each vertex.
In~\cite{wong2015grover}, the authors mainly investigate the effect of extra
self-loops on Grover's algorithm when formulated as search for a marked vertex on complete graphs.
They find that adding self-loops can either slow down or boost the success probability
by choosing different coin operators.
On the other hand, three-state quantum walks on a line have been investigated exhaustively.
Most notably, if the walker of a three-state quantum walk is initialized at one site,
it will be trapped with large probability near the origin after walking enough steps~\cite{inui2005one, falkner2014weak}.
This phenomenon is previously found in quantum walks on
square lattices~\cite{inui2004localization} and is called localization.
Researches show that the localization effect happens with a broad family of coin
operators in three-state quantum walks
~\cite{vstefavnak2012continuous,vstefavnak2014stability,vstefavnak2014limit}.
Moreover, a weak limit theorem is recently derived in~\cite{falkner2014weak, machida2015limit}
for arbitrary coin initial state and coin operator.
However, the properties of LQWs, such as localization and spread behavior,
are still open. In this paper, we give a in-depth study the LQWs on a line.
Since the lackadaisical model is more complicated than the standard one,
we could expect more intrinsic characteristics.

The rest of this paper is organized as follows.
In Sec.~\ref{sec:definition}, we give formal definitions of LQWs
and describe the Fourier transformation method which is often used in analyzing quantum walks.
In Sec.~\ref{sec:localization}, we provide a mathematical framework for the walker's localization probability
on the time limit.
In Sec.~\ref{sec:velocity}, we find the explicit forms to compute the velocities of
the left- and right-travelling peaks appeared in the walker's probability distribution.
And in Sec.~\ref{sec:weak-limit}, we obtain a long time approximation for the entire probability density function
and prove that all LQWs spread ballistically.
Finally, we conclude in Sec.~\ref{sec:conclusion}.

\section{Definitions}\label{sec:definition}

\subsection{Lackadaisical quantum walks}

\begin{figure}
  \centering
  \includegraphics[width=0.4\textwidth]{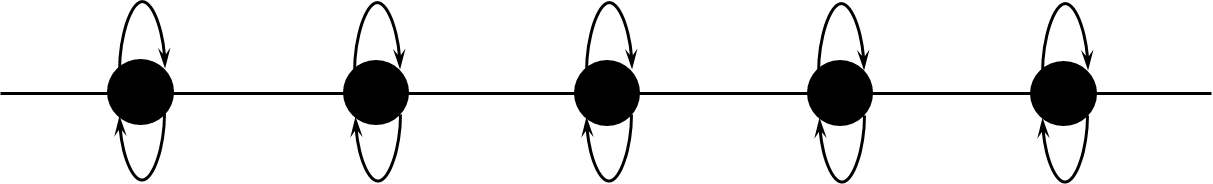}
  \caption{An illustrative example of an infinite line with $2$ self-loops attached to each vertex.}\label{fig:self-loops}
\end{figure}

In this paper, a LQW is defined to be a quantum walk
on an infinite line with $\tau$ self-loops attached to each vertex.
An illustrative example is given in Fig.~\ref{fig:self-loops}, in which each vertex has 2 additional self-nodes.
We term the number of self-loops $\tau$ as the \textit{laziness factor}.
If $\tau=0$, it is the standard quantum walk (also called the Hadamard walk).
In this paper we consider $\tau>0$.
It's obvious that in lazy random walks, the greater the $\tau$ is, the more the walker prefers to stay.
The total system of a LQW with laziness factor $\tau$
is given by $ \mathcal{H} = \mathcal{H}_P \otimes \mathcal{H}_C$,
where $\mathcal{H}_P$ is the position space defined as
$$ \mathcal{H}_P = \textrm{Span}\{\ket{n}, n \in \mathbb{Z}\},$$
and $\mathcal{H}_C$ is the coin space.
In each step, the walker has $\Delta$ ($\Delta = \tau + 2$) choices - 
it can move to the left, or the right,
or just stay in current position via a self-loop.
For each of these options, we assign a standard basis of the coin space $\mathcal{H}_C$.
Thus $\mathcal{H}_C$ is defined as 
$$\mathcal{H}_C = \mathbb{C}^\Delta = \textrm{Span}\{\ket{1}, \;\ket{2}, \;\cdots, \;\ket{\Delta}\}.   $$
A single step of quantum walk is given by
$U = S \cdot (\mathbb{I}_P \otimes C)$
where $S$ is the position shift operator, $\mathbb{I}_P$ is the identity of 
$\mathcal{H}_P$ and $C$ is the coin flip operator.
For LQWs, the position shift operator $S$ is
\begin{eqnarray*}
  S &=& \sum_{n \in \mathbb{Z}}\Big\{
                  \ket{n-1}\bra{n} \otimes \ket{1}\bra{1} + \ket{n+1}\bra{n} \otimes \ket{2}\bra{2} \\
  ~ &~& \quad\quad + \sum_{j=3}^{\Delta} \ket{n}\bra{n} \otimes \ket{j}\bra{j} \Big\}.
\end{eqnarray*}
For the coin operator $C$, a common choice is the Grover operator $G$, which is defined as
\begin{equation}\label{eq:coin-operator}
    \operator{G} = \frac{1}{\Delta}
        \begin{pmatrix}
          -\tau   & 2       & 2       & \cdots  & 2   \\
          2       & -\tau   & 2       & \cdots  & 2   \\
          2       & 2       & -\tau   & \cdots  & 2   \\
          \vdots  & \vdots  & \vdots  & \vdots  & \vdots   \\
          2       & 2       & \cdots  & 2       & -\tau
        \end{pmatrix}.
\end{equation}
Let
$\ket{\Psi(t, n)} = [\psi_1(t, n), \psi_2(t, n), \cdots, \psi_\Delta(t, n)]^\dag \in \mathcal{H}_C$
be the probability amplitude of the walker at position $n$ at time $t$,
then the system state can be expressed by
$$
\ket{\Psi(t)} = \sum_{n \in \mathbb{Z}} \ket{n}\otimes \ket{\Psi(t, n)}.
$$
$\ket{\Psi(t)}$ can be obtained by applying $U$ to the initial state $\ket{\Psi(0)}$ for $t$ times,
i.e. $\ket{\Psi(t)} = U^t \ket{\Psi(0)}$.
The walker $X_t$ can be found at position $n$ at time $t$ with probability
\begin{equation}\label{eq:calculate-prob}
  \mathbb{P}(X_t = n)  = \langle\Psi(t,n) \vert \Psi(t,n)\rangle = \sum_{j=1}^{\Delta}\vert \psi_j(t,n) \vert^2.
\end{equation}
Expanding $\ket{\Psi(t+1)} = U \ket{\Psi(t)}$ in terms of $\ket{\Psi(t, n)}$,
we obtain the master equation for the walker at position $n$
\begin{eqnarray}\label{eq:recurrence}
  \ket{\Psi(t+1, n)} &=& \quad G_1\ket{\Psi(t, n+1)} + G_2\ket{\Psi(t, n-1)} \nonumber\\
  ~ &~& + \sum_{j=3}^{\Delta}G_j\ket{\Psi(t, n)},
\end{eqnarray}
where $G_j = \sum_{k=1}^{\Delta}G_{j,k}\ket{j}\bra{k}, j = 1, 2, \cdots, \Delta$, and
$G_{j,k}$ are the elements of $G$ defined in Eq.~\ref{eq:coin-operator}.

\subsection{Fourier analysis}
Eq.~\ref{eq:recurrence} can be solved by Fourier transformation on the system state
\begin{equation}\label{eq:fourier}
\ket{\tilde{\Psi}(t, k)} = \sum_{n \in \mathbb{Z}} e^{-ikn} \ket{\Psi(t, n)},\;\; k \in (-\pi, \pi].
\end{equation}
From now on, a tilde indicates quantities with a $k$ dependence.
The inverse Fourier transform is
\begin{equation}\label{eq:inverse-fourier}
\ket{\Psi(t, n)} = \int_{-\pi}^{\pi}\frac{dk}{2\pi} e^{ink} \ket{\tilde{\Psi}(t, k)}.
\end{equation}
Substituting Eq.~\ref{eq:fourier} to Eq.~\ref{eq:recurrence} yields the master equation in the Fourier space
\begin{equation}\label{eq:fourier-recurrence}
  \tilde{\Psi}(t+1, k) =
      \underbrace{ \big[ G_1e^{ik} + G_2e^{-ik} + \sum_{j=3}^{\Delta}G_j\big]}_{\tilde{U}_k} \tilde{\Psi}(t, k).
\end{equation}
Let $\kappa = e^{ik}$, $\tilde{U}_k$ has the form of
\begin{equation*}
    \tilde{U}_k = \frac{1}{\Delta}
        \begin{pmatrix}
          -\tau\kappa & 2\kappa       & 2\kappa       & \cdots  & 2\kappa   \\
          2/\kappa    & -\tau/\kappa  & 2/\kappa  & \cdots  & 2/\kappa   \\
          2       & 2       & -\tau   & \cdots  & 2   \\
          \cdots  & \cdots  & \cdots  & \cdots  & \cdots   \\
          2       & 2       & \cdots  & 2       & -\tau
        \end{pmatrix}.
\end{equation*}
Since $\tilde{U}_k$ is unitary, its eigenvalues have the forms of $\lambda_j = e^{i\omega_j}$.
We denote $\ket{\lambda_j}$ as the corresponding eigenvectors.
After some calculation, we get explicit forms of the eigenvalues
\begin{equation*}
  \omega_j =
        \left\{ \begin{array}{ll}
          \theta, & j=1,\\
          -\theta, & j=2,\\
          0, & j=3, \\
          \pi, & j \geq 4, \\
\end{array} \right.
\end{equation*}
where $\theta$ satisfies
\begin{eqnarray*}
  \cos\theta &=& -\frac{\tau\cos k + 2}{\tau + 2}, \\
  \sin\theta &=& \frac{\sqrt{\tau(1 - \cos k)(\tau + 4 + \tau\cos k)}}{\tau + 2}.
\end{eqnarray*}
The corresponding eigenvectors are
\begin{equation*}
\ket{\lambda_j} = \sqrt{N_j}
\begin{pmatrix}
  \frac{1}{1 + e^{i(\omega_j - k)}} \vspace{0.1in}\\
  \frac{1}{1 + e^{i(\omega_j + k)}} \vspace{0.1in}\\
  \frac{1}{1 + e^{i\omega_j}} \vspace{0.1in}\\
  \vdots \\
  \vdots \vspace{0.1in} \\
  \frac{1}{1 + e^{i\omega_j}}
\end{pmatrix}, \;\mbox{for}\; j = 1, 2, 3;
\end{equation*}
\begin{equation}\label{eq:eigenvectors}
\ket{\lambda_j} = \frac{1}{\sqrt{2}}
\begin{pmatrix}
  0 \\
  0 \\
  -1 \\
  0   \\
  \vdots \\
  0   \\
  1 \\
  0   \\
  \vdots
\end{pmatrix}
\begin{matrix}
  ~ \\
  ~ \\
  ~ \\
  ~ \\
  ~ \\
  ~ \\
  \leftarrow \mbox{$j$-th row} \\
  ~ \\
  ~
\end{matrix}, \;\forall j \geq 4,
\end{equation}
where $N_j$ is the corresponding normalization factor.
Putting $\tilde{U}_k$ in its eigenbasis, we can rewrite Eq.~\ref{eq:fourier-recurrence} as
\begin{eqnarray}\label{eq:evolution-equation}
  \tilde{\Psi}(t, k)  &=& \tilde{U}_k \tilde{\Psi}(t, k) = \tilde{U}_k^t \tilde{\Psi}(0, k) \nonumber\\
  ~                   &=& \sum_{j=1}^{\Delta}\lambda_j^t\ketbra{\lambda_j} \tilde{\Psi}(0, k),
\end{eqnarray}
where $\tilde{\Psi}(0, k)$ is the Fourier transformed initial state.

In this paper, we assume the walker always starts at position $0$ and the initial coin state satisfies
$\ket{\Psi(0, 0)} = \alpha\ket{1} + \beta\ket{2}$, where $\alpha, \beta \in \mathbb{C}$,  $\alpha^2 + \beta^2 = 1$.
This assumption is quite reasonable if
we want to have a same initial state for walks with different $\tau$.
We guarantee the quantum walker starts with its coin state in superposition of only left and right bases.
Therefore, the system's initial state can be formulated as
\begin{equation}\label{eq:init-state}
\Psi(0) = \sum_{n \in \mathbb{Z}} \delta_{n,0} \ket{0} \otimes \big[\alpha\ket{1} + \beta\ket{2}\big],
\end{equation}
where $\delta_{n,0}$ is the Kronecker function.
By Eq.~\ref{eq:fourier} the Fourier transformed system's initial state becomes
\begin{equation}\label{eq:dfted-init-state}
\tilde{\Psi}(0, k) = [\alpha, \beta, 0, \cdots, 0]^\dag, \quad \forall k \in (-\pi, \pi].
\end{equation}

\section{Probability at Origin}\label{sec:localization}

\begin{figure}
  \centering
  \includegraphics[width=0.45\textwidth]{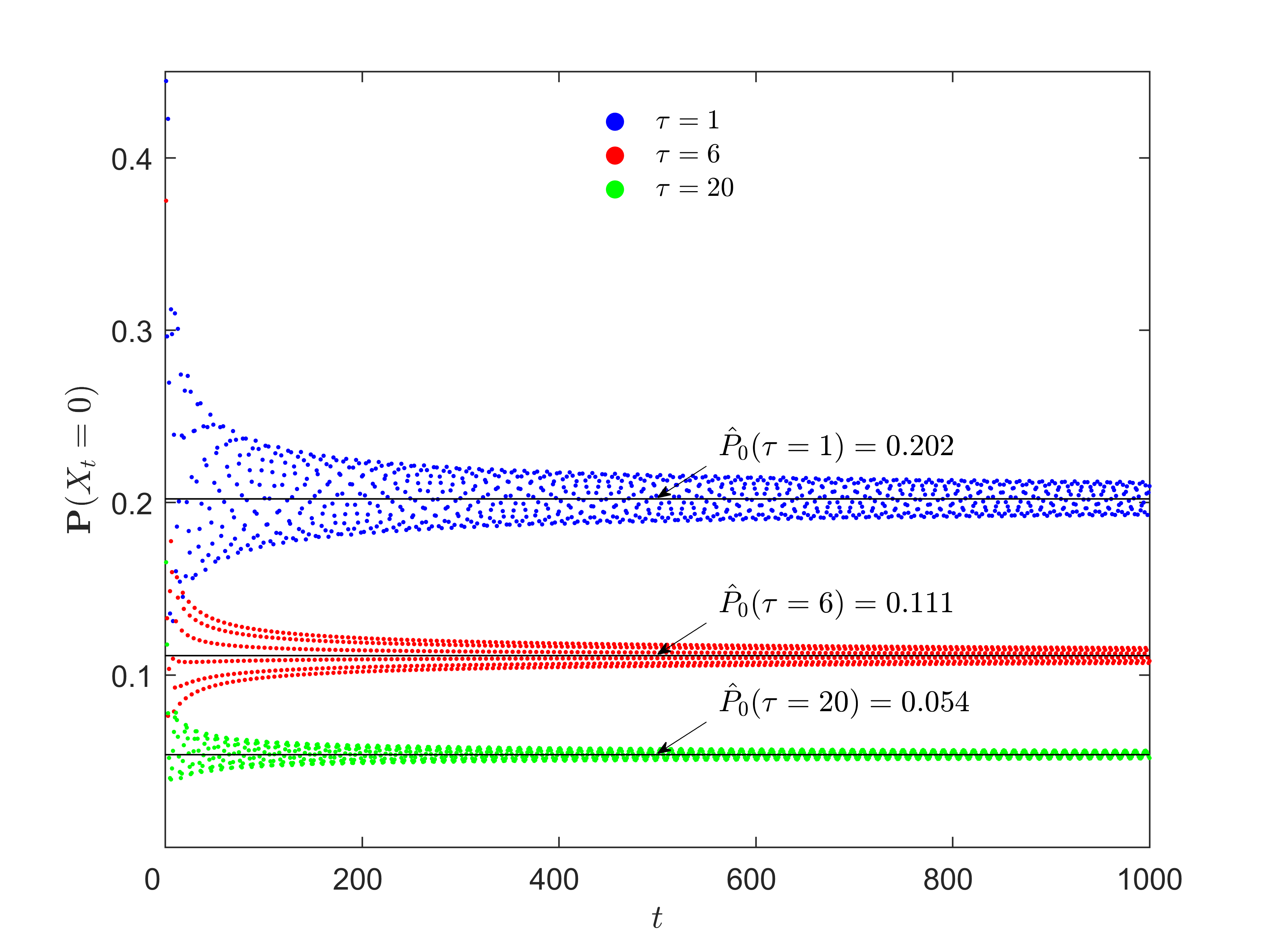}
  \caption{Numerical probabilities of finding the walker at the origin as a function of walking steps
          for LQWs with various laziness factors $\tau=1$ (blue dots),  $\tau=6$ (red dots),
          and $\tau=20$ (green dots). These factors are carefully chosen to show different oscillating behaviors.
          The initial coin state is $\alpha=\frac{1}{2}$, $\beta=\frac{i}{2}$ for all walks.
          The horizontal lines are the corresponding theoretical localization probabilities obtained
          by Eq.~\ref{eq:localization-prob}.
          }\label{fig:origin-probability}
\end{figure}

In this section, we focus on the localization phenomenon on LQWs.
To determine whether the localization will occur at the origin, we need to calculate
$\lim_{t \rightarrow \infty} \mathbb{P}(X_t = 0)$.
Let the probability be $\hat{P}_0$, where we use a caret ($\hat{~}$) to indicate
the asymptotic limit of $t$.
\begin{theorem}\label{thm:localization-prob}
  For a LQW with laziness factor $\tau$,
  if the walker starts with the state given by Eq.~\ref{eq:init-state},
  the asymptotic limit of the probability of the walker at origin is
  \begin{equation}\label{eq:localization-prob}
    \hat{P}_0 = 2\cdot\frac{\tau + 4 - 2\sqrt{2\tau+4}}{\tau^2}.
  \end{equation}
\end{theorem}
It's obvious from the theorem that, if the walker starts on a superposition of only left and right directions,
the localization probability of the walker is independent on the coin initial state, and is
totally dominated by the laziness factor $\tau$.
When $\tau=1$, we get $\hat{P}_0 = 2(5-2\sqrt{6})$.
This result coincides with Eq.~15 derived in~\cite{falkner2014weak} when $\beta=0$.
We perform numerical simulations and the conclusions are summarized in Fig.~\ref{fig:origin-probability}.
The figure manifests the probabilities $\mathbb{P}(X_t = 0)$ oscillate around their corresponding
theoretical limiting values $\hat{P}_0$
for $\tau=1$,  $\tau=6$, and $\tau=20$.
It is clearly that, for different laziness factors,
the probability at the origin oscillates periodically with different patterns.
These oscillations clearly exhibit tendencies to converge, indicating that
the walker does have a non-zero probability to be localized.
Furthermore, we observe that the larger the $\tau$ is, the faster the probabilities converge.
\begin{proof}
By Eq.~\ref{eq:calculate-prob}, we have
\begin{equation}\label{eq:origin-prob-expression}
  \hat{P}_0 \equiv \lim_{t \rightarrow \infty} \mathbb{P}(X_t = n)
            = \lim_{t \rightarrow \infty} \langle\Psi(t,0) \vert \Psi(t,0)\rangle.
\end{equation}
To obtain $\hat{P}_0$, we have to calculate
$\lim_{t \rightarrow \infty} \ket{\Psi(t,0)}$.
Substitute Eq.~\ref{eq:evolution-equation} into Eq.~\ref{eq:inverse-fourier} and let $n = 0$,
we derive the explicit form for $\ket{\Psi(t,0)}$
\begin{equation*}
  \ket{\Psi(t, 0)}  = \int_{-\pi}^{\pi}\frac{dk}{2\pi} \tilde{\Psi}(t, k)
            = \int_{-\pi}^{\pi}\frac{dk}{2\pi} \sum_{j=1}^{\Delta}
                \lambda_j^t\ket{\lambda_j}\langle\lambda_j\vert\tilde{\Psi}(0, k)\rangle.
\end{equation*}
From \textbf{Lemma 1} in~\cite{lyu2015localization}, we know that
the contributions to $\ket{\Psi(t, 0)}$ from items with $j=1, 2$ in the above equation
are negligible when $t$ approaches infinity.
As a result, $\ket{\Psi(t, 0)}$ is totally determined by the integrals with $j\geq 3$.
Since $\lambda_3 = 1$, and $\forall j \geq 4,\lambda_j = -1$,
we can further simplify the equation above by substituting into these constant eigenvalues.
The final expression is shown in Eq.~\ref{eq:origin-amplitude-2}.
In this equation, only $\ket{\lambda_3}$ is a function of $k$,
while for all $j \geq 4, \ket{\lambda_j}$ and
$\tilde{\Psi}(0, k)$ are independent on $k$ according to Eq.~\ref{eq:eigenvectors} and~\ref{eq:dfted-init-state} respectively.
Actually, Eq.~\ref{eq:origin-amplitude-2} can be understood as a series of linear maps from the
initial state $\tilde{\Psi}(0, k)$ to $\Psi(t, 0)$.
The linear maps are represented by a set of transformation matrices $F_j$ defined as
\begin{widetext}
\begin{eqnarray}\label{eq:origin-amplitude-2}
  \lim_{t \rightarrow \infty} \ket{\Psi(t, 0)}  &\sim& \int_{-\pi}^{\pi}\frac{dk}{2\pi}
                \Big[ \ketbra{\lambda_3} + \sum_{j=4}^{\Delta}(-1)^t\ketbra{\lambda_j} \Big] \ket{\tilde{\Psi}(0, k)} \nonumber \\
  ~           &=& \Bigg[   \int_{-\pi}^{\pi}\frac{dk}{2\pi}\ketbra{\lambda_3}
                        + \sum_{j=4}^{\Delta}(-1)^t \int_{-\pi}^{\pi}\frac{dk}{2\pi}\ketbra{\lambda_j} \Bigg] \ket{\tilde{\Psi}(0, k)},
\end{eqnarray}
\end{widetext}
\begin{eqnarray}
  F_3 &=& \int_{-\pi}^{\pi}\frac{dk}{2\pi} \ketbra{\lambda_3},\label{eq:f_3} \\
  F_j &=& \int_{-\pi}^{\pi}\frac{dk}{2\pi} \ketbra{\lambda_j} = \ketbra{\lambda_j}, \; \forall j \geq 4.\label{eq:f_j}
\end{eqnarray}
The matrices $F_j$ capture all information about the walker's behavior at its original position
when $t \rightarrow \infty$. The existence of localization is directly related to the system's initial state
via the matrices $F_j$.
For all $j \geq 4$, $F_j$ is independent on $k$, so it is a constant matrix and can be easily calculated.
The exact form of $F_3$ can be obtained by exploiting the eigenvector $\ket{\lambda_3}$.
Let $\kappa_1 = \frac{2}{1 + e^{-ik}}$, $\kappa_2 = \frac{2}{1 + e^{ik}}$,
it's obvious that $\kappa_1^\dag = \kappa_2$ and $\kappa_2^\dag = \kappa_1$.
Substitute $\omega_3 = 0$ into Eq.~\ref{eq:eigenvectors} for $j=3$,
we get the explicit form of $\ket{\lambda_3}$
\begin{eqnarray*}
  \ket{\lambda_3} &=& \sqrt{N_3}\big[\kappa_1,\; \kappa_2,\; 1,\; \cdots,\; 1\big]^\dag,
\end{eqnarray*}
where
$
N_3 = \frac{1}{\kappa_1\kappa_1^\dag + \kappa_2\kappa_2^\dag + \tau}
    = \frac{\tau + 4 + \tau\cos k}{1 + \cos k}
$ is the normalization factor.
Then
\begin{equation*}
  F_3  = \int_{-\pi}^{\pi}\frac{dk}{2\pi}N_3
      \begin{pmatrix}
        \kappa_1\kappa_2  & \kappa_1\kappa_1   & \kappa_1  & \cdots  & \kappa_1 \\
        \kappa_2\kappa_2  & \kappa_2\kappa_1   & \kappa_2  & \cdots  & \kappa_2 \\
        \kappa_2          &         \kappa_1   & 1         & \cdots  & 1        \\
        \vdots            &         \vdots     & \vdots    & \vdots  & \vdots   \\
        \kappa_2          &         \kappa_1   & 1         & \cdots  & 1
      \end{pmatrix}.
\end{equation*}
Define
$\Theta_1 = \frac{1}{\tau} - \frac{\sqrt{2\tau + 4}}{\tau(\tau + 2)}$,
$\Theta_2 = \frac{\sqrt{2\tau + 4}}{2\tau + 4}$, and
$\Theta_3 = \frac{2}{\tau} - \frac{(\tau+4)\sqrt{2\tau + 4}}{2\tau(\tau + 2)}$,
we can show after some tedious calculations
\begin{eqnarray*}
  &~&\int_{-\pi}^{\pi}\frac{dk}{2\pi}N_3 =
  \int_{-\pi}^{\pi}\frac{dk}{2\pi}N_3\kappa_1 =
  \int_{-\pi}^{\pi}\frac{dk}{2\pi}N_3\kappa_2 = \Theta_1, \\
  &~&\int_{-\pi}^{\pi}\frac{dk}{2\pi}N_3\kappa_1\kappa_2 =
  \int_{-\pi}^{\pi}\frac{dk}{2\pi}N_3\kappa_2\kappa_1 = \Theta_2,\\
  &~&\int_{-\pi}^{\pi}\frac{dk}{2\pi}N_3\kappa_1\kappa_1 =
  \int_{-\pi}^{\pi}\frac{dk}{2\pi}N_3\kappa_2\kappa_2 = \Theta_3.
\end{eqnarray*}
Thus the explicit form of $F_3$ is
\begin{equation*}
  F_3 = \begin{pmatrix}
          \Theta_2 & \Theta_3   & \Theta_1  & \cdots  & \Theta_1 \\
          \Theta_3 & \Theta_2   & \Theta_1  & \cdots  & \Theta_1 \\
          \Theta_1 & \Theta_1   & \Theta_1  & \cdots  & \Theta_1 \\
          \vdots   & \vdots     & \vdots    & \vdots  & \vdots   \\
          \Theta_1 & \Theta_1   & \Theta_1  & \cdots  & \Theta_1
        \end{pmatrix}.
\end{equation*}
Substitute $F_j$ into Eq.~\ref{eq:origin-amplitude-2}, we get
\begin{equation}\label{eq:final-state}
\lim_{t \rightarrow \infty} \ket{\Psi(t, 0)} = \Big[ F_3 + \sum_{j=4}^{\Delta}(-1)^t F_j \Big] \ket{\tilde{\Psi}(0, k)},
\end{equation}
where $\ket{\tilde{\Psi}(0, k)}$ is given in Eq.~\ref{eq:dfted-init-state}.
Let $\lim_{t \rightarrow \infty} \ket{\Psi(t, 0)} = [\phi_1, \cdots, \phi_\Delta]^\dag$,
as $F_j$ has no impact on the first two components of $\ket{\tilde{\Psi}(0, k)}$ for all $j \geq 4$,
Eq.~\ref{eq:final-state} can be easily solved:
\begin{eqnarray*}
  \phi_1 &=& \Theta_2\alpha + \Theta_3\beta, \\
  \phi_2 &=& \Theta_3\alpha + \Theta_2\beta, \\
  \phi_j &=& \Theta_1(\alpha + \beta), \; \forall j \geq 3.
\end{eqnarray*}
The asymptotic limit of the probability at origin $\hat{P}_0$ can be obtained now
\begin{equation*}
  \hat{P}_0 = \vert\phi_1\vert^2 + \vert\phi_2\vert^2 + \tau\vert\phi_3\vert^2
            = 2\cdot\frac{\tau + 4 - 2\sqrt{2\tau+4}}{\tau^2}.
\end{equation*}
\end{proof}

Though Theorem~\ref{thm:localization-prob} only considers localization probabilities
for a special class of initial states (given in Eq.~\ref{eq:init-state}),
we should point out that we are able to calculate the localization probability
by Eq.~\ref{eq:origin-prob-expression} and~\ref{eq:final-state} for arbitrary system's initial state
that satisfies
\begin{equation*}
\Psi(0) = \sum_{n \in \mathbb{Z}} \delta_{n,0} \ket{0} \otimes \sum_{j=1}^{\Delta} \alpha_j\ket{j},
\end{equation*}
where $\alpha_j \in \mathbb{C}$, and $\sum_{j=1}^{\Delta} \vert\alpha_j\vert^2 = 1$.

\section{Peak Velocity}\label{sec:velocity}

In this section, we determine the peak velocity at which LQWs spread on the line.
The analytical method we use here is first described in~\cite{vstefavnak2012continuous}.
From their arguments, we know that the peak velocity is given by the first order of the stationary
points of the phase
\begin{equation*}
  \tilde{\omega}_j \equiv \omega_j  - \frac{n}{t}k.
\end{equation*}
The stationary point of the second order of $\tilde{\omega}_j$ corresponds to the solution of $k$.
We should notice that both the first and second derivatives of $\tilde{\omega}_j$ with respect to $k$ vanish.
Therefore in order to obtain the peak velocity, we need to solve equations
\begin{eqnarray}\label{eq:first-second-order}
  \frac{d\tilde{\omega}_j}{dk} &=& \frac{d\omega_j}{dk} - \frac{n}{t} = 0,  \nonumber\\
  \frac{d^2\tilde{\omega}_j}{dk^2} &=& \frac{d^2\omega_j}{dk^2} = 0.
\end{eqnarray}
Assume $k_0$ is the solution of the second equation in Eq.~\ref{eq:first-second-order}, then
by the first equation we obtain the position of the peak after $t$ steps
$$
n = \frac{d\omega_j}{dk}\Big\vert_{k_0} t.
$$
The peak propagates with a constant velocity $\frac{d\omega_j}{dk}\big\vert_{k_0}$.

\begin{figure}
  \centering
  \includegraphics[width=0.45\textwidth]{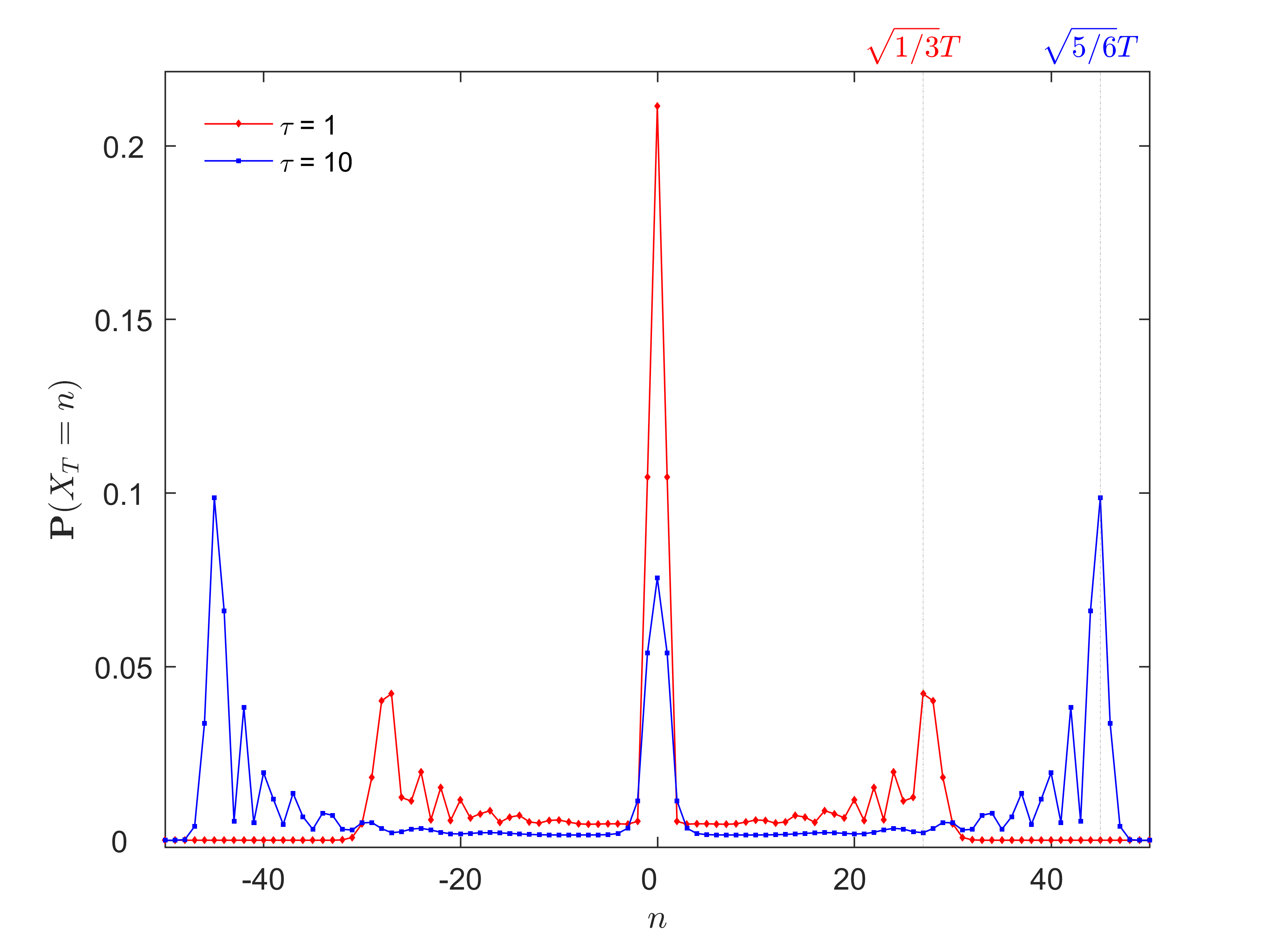}
  \caption{ Probability distributions of LQWs after $T = 50$ steps,
            for various laziness factors $\tau$.
            The coin initial state is $\alpha=1/\sqrt{2}$, $\beta=i/\sqrt{2}$,
            This state will give a symmetric walk.
            We can easily identify three dominant peaks in each probability distribution.
            We visualize the right peaks $\sqrt{1/3}T \approx 29$, $\sqrt{5/6}T \approx 46$ in grid lines
            for $\tau=1$ and $\tau=10$. These theoretical peaks coincide with the positions of peaks
            obtained from numerical simulations.
             }\label{fig:peak-velocity}
\end{figure}

Now we show the peak velocities for LQWs on a line.
As the phases $\omega_j$ are constant for all $j \geq 3$, we immediately know that their corresponding
peak velocities are $v_S = 0$. This is easy to understand as the constant phases result in
the central peak of the probability distribution staying    .
Thus the velocities of left and right travelling peaks are dominated by $\omega_{1,2}$.
We find the equations in Eq.~\ref{eq:first-second-order} can be solved
by investigating $\omega_1$ and $\omega_2$
\begin{eqnarray}
  \frac{d\omega_{1,2}}{dk} &=& \pm \frac{\tau\sin k}{\sqrt{\tau(1-\cos k)(\tau\cos k + \tau + 4)}},\label{eq:first-derivate} \\
  \frac{d^2\omega_{1,2}}{dk^2} &=& \pm 2\sqrt{\frac{\tau(1 - \cos k)}{ (\tau\cos k + \tau + 4)^3}}. \nonumber
\end{eqnarray}
In $k \in (-\pi, \pi]$, $d^2\omega_{1,2}/dk^2=0$ has a solution when $k_0 = 0$.
Evaluating $\omega_{1,2}/dk$ at $k_0$, we get the peak velocities of the left and right traveling probabilities
\begin{eqnarray}
  v_R &=& \lim_{k \rightarrow 0^{+}} \frac{d \omega_2}{dk} = \sqrt{\frac{\tau}{\tau + 2}}, \label{eq:right-velocity} \\
  v_L &=& \lim_{k \rightarrow 0^{+}} \frac{d \omega_1}{dk} = - \sqrt{\frac{\tau}{\tau + 2}}.\label{eq:left-velocity}
\end{eqnarray}
When the laziness factor satisfies $\tau=1$,
we recover the results presented in~\cite{vstefavnak2012continuous}.
As an illustrative example, we plot the walker's probability distribution of the LQW
whose laziness factor is $10$ in Fig.~\ref{fig:peak-velocity}.
The probability distribution contains three dominant peaks,
the left and right travelling peaks are given by the peak velocities $v_L$ and $v_R$ respectively.

From Eq.~\ref{eq:right-velocity} (Eq.~\ref{eq:left-velocity}) we can see that as the laziness factor increases,
the right (left) peak velocity also becomes larger.
In this sense, we can control the spread behavior the quantum walker
and achieve faster spreading than the standard quantum walks.
In~\cite{vstefavnak2012continuous}, the authors offer a different way to
control the spread behavior of the walker by tuning the parameter $\rho$ of
the generalized Grover coin operator (see Eq.~14 in their paper), the underlying
quantum walks are still three-state quantum walks.
While in our paper we actually propose a \textit{multi-}state quantum walk scheme by
introducing different number of self-loops to each vertex,
the spread behavior of the walker can be controlled by tuning the laziness factor $\tau$,
the underlying coin operator is always Grover operator.
In the extreme case we have
$
\tau \rightarrow \infty \Rightarrow \vert v_{R,L} \vert = 1.
$
This indicates that if there is infinite self-loops in each vertex,
the quantum walker will propagate on the line with constant speed $1$.
This can be explained by investigating the coin operator $G$ defined in Eq.~\ref{eq:coin-operator}.
When $\tau \rightarrow \infty$, $G$ satisfies
\begin{eqnarray*}
  \lim_{\tau \rightarrow \infty} G &\equiv& \lim_{\tau \rightarrow \infty} 2\ket{\psi}\bra{\psi} - \mathbb{I}_C \\
  ~ &=& \lim_{\tau \rightarrow \infty} \frac{2}{\tau+2}\sum_{j=1}\sum_{k=1}\ket{j}\bra{k} - \mathbb{I}_C \sim -\mathbb{I}_C,
\end{eqnarray*}
where $\mathbb{I}_C$ is the identity of $\mathcal{H}_C$.
That is, when $\tau \rightarrow \infty$, the coin operator $G$ approximates to $-\mathbb{I}_C$, 
which results in a trivial quantum walk.
This fast spread behavior of LQWs is striking different from lazy random walks, 
in which the additional self-loops will slow down the spread speed.
In the extreme case where $\tau \rightarrow \infty$, the classical walker will localize
in the origin and never spread. 
\section{Weak Limit}\label{sec:weak-limit}

In this section, we present a weak limit distribution for the rescaled LQW $X_t/t$ as $t \rightarrow \infty$.
It expresses an asymptotic behavior of the walk after long enough time.
The limit distribution is composed of a Dirac $\delta$-function
related to the localization probability calculated in Sec.~\ref{sec:localization} and
a continuous function with a compact support whose domain is given by
the peak velocities given in Sec.~\ref{sec:velocity}.
We also prove that LQWs spread ballistically.
The analytical method we use in this section is first proposed in~\cite{grimmett2004weak}
and we mainly follow the proof procedure outlined in~\cite{machida2015limit}.
What's more, one should keep in mind that in this paper we only consider a special 
class of system's initial states defined in Eq.~\ref{eq:init-state}.
\begin{theorem}\label{thm:weak-limit}
  For any real number $x$, we have
  \begin{equation*}
    \lim_{t \rightarrow \infty} \mathbb{P}(\frac{X_t}{t} \geq x) =
      \int_{-\infty}^{x}dy \Big\{ \delta_{0,y}\hat{P} + f(y)I_{(-\Omega, \Omega)}(y) \Big\},
  \end{equation*}
  where
  \begin{itemize}
    \item $\delta_{0,y}$ is the Dirac $\delta$-function at the origin,
    \item $\hat{P}$ is the sum of localization probabilities in all positions and satisfies
      \begin{equation*}
        \hat{P} = \frac{\sqrt{2\tau + 4}}{2\tau + 4}
                + 2\Big\{\frac{2}{\tau} - \frac{(\tau+4)\sqrt{2\tau + 4}}{2\tau(\tau + 2)}\Big\}\mathfrak{R}(\alpha^\dag\beta),
      \end{equation*}
    \item $f(y)$ is the weak limit density function defined in Eq.~\ref{eq:f-x},
    \item $\Omega = \sqrt{\frac{\tau}{\tau+2}}$ is the bound of the compact support domain, and
    \item $I_{\Gamma}(y)$ is the compact support function whose domain is $\Gamma$ and defined as
      \begin{equation*}
        I_{\Gamma}(y) = \left\{ \begin{array}{ll}
                        1, & y \in \Gamma, \\
                        0, & y \notin \Gamma.
                        \end{array} \right.
      \end{equation*}
  \end{itemize}
\end{theorem}
From the theorem we can see that the limit density function rescaled by time has a compact support
and its domain $(-\Omega, \Omega)$ is totally determined by the walker's travelling peak velocities.
A weak limit theorem of three-state walks is presented in~\cite{machida2015limit}.
Our results are the same as theirs when we let $\tau=1$ and set the parameters $c=-1/3$, $s=2\sqrt{2}/3$, $\beta=0$
in Theorem 2 of their paper.
One should note the difference between $\hat{P}$ and $\hat{P}_0$ (the localization probability at the origin)
studied in Sec.~\ref{sec:localization}.
Actually, $\hat{P}$ is the sum of localization probabilities in all positions,
i.e., $\hat{P} = \sum_{n\in\mathbb{Z}} \hat{P}_n$.
We are unable to derive an analytic expression for $\hat{P}_n$ for $n \neq 0$,
but luckily we can still calculate $\hat{P}$.
 \begin{proof}
The $r$-th moment of the quantum walker's probability distribution can be calculated as 
\begin{eqnarray*}
  \mathbb{E}(X_t^r) &=& \sum_{n \in \mathbb{Z}} n^r \mathbb{P}(X_t = n)  \\
  ~                 &=& \int_{-\pi}^{\pi}\frac{dk}{2\pi}
                        \langle \tilde{\Psi}(t, k)\vert\Big( D^r\ket{\tilde{\Psi}(t, k)} \Big)  \\
  ~                 &=& (t)_r \int_{-\pi}^{\pi}\frac{dk}{2\pi}
                        \sum_{j=1}^{\Delta} \Big( i\frac{\lambda_j^{\prime}}{\lambda_j}\Big)^r
                        \Big\vert \langle \lambda_j \vert \tilde{\Psi}(0, k) \rangle  \Big\vert^2 + O(t^{r-1}),
\end{eqnarray*}
where $D = i(d/dk)$ and $(t)_r = t(t-1)\cdots(t-r+1)$.
To have $X_t$ spatially rescaled by time, we divide both sides of the above equation by $t^r$ and
take a limit on $t$
\begin{eqnarray}\label{eq:moment-limit}
  \lim_{t \rightarrow \infty} \mathbb{E}\Big[\Big(\frac{X_t}{t}\Big)^r \Big] &=&
      \quad\sum_{j=1}^{2}\int_{-\pi}^{\pi}\frac{dk}{2\pi}
      \Big( i\frac{\lambda_j^{\prime}}{\lambda_j}\Big)^r
      \Big\vert \langle \lambda_j \vert \tilde{\Psi}(0, k) \rangle  \Big\vert^2 \nonumber\\
  ~ &~& + \sum_{j=3}^{\Delta}\int_{-\pi}^{\pi}\frac{dk}{2\pi}
      \Big\vert \langle \lambda_j \vert \tilde{\Psi}(0, k) \rangle  \Big\vert^2.
\end{eqnarray}
As $F_j$ has no impact on the first two components of $\ket{\tilde{\Psi}(0, k)}$ for all $j \geq 4$,
the second term in Eq.~\ref{eq:moment-limit} can be easily calculated by making use of
the transformation matrix $F_3$ defined in Eq.~\ref{eq:f_3}
\begin{eqnarray}\label{eq:part1}
    \sum_{j=3}^{\Delta}\int_{-\pi}^{\pi}\frac{dk}{2\pi}
    \Big\vert \langle \lambda_j \vert \tilde{\Psi}(0, k) \rangle  \Big\vert^2
  &=&
    \int_{-\pi}^{\pi}\frac{dk}{2\pi}\Big\vert \langle \lambda_3 \vert \tilde{\Psi}(0, k) \rangle  \Big\vert^2\nonumber\\
  ~ &=&
    \bra{\tilde{\Psi}(0, k)}\cdot F_3\cdot \ket{\tilde{\Psi}(0, k)} \nonumber\\
  ~ &=&
    \Theta_2 + 2\Theta_3\mathfrak{R}(\alpha^\dag\beta).
\end{eqnarray}
Then we calculate the first term.
As $\lambda_j = e^{i\omega_j}$, we can get the derivation of $\lambda_j$
using the expressions for $d\omega_j/dk$ obtained in Eq.~\ref{eq:first-derivate}
for $j = 1, 2$
$$
  i\frac{\lambda_j^\prime}{\lambda_j}
= -\frac{d\omega_j}{dk}
= (-1)^{j-1}\frac{\tau\sin k}{\sqrt{\tau(1-\cos k)(\tau\cos k + \tau + 4)}}.
$$
Putting $i\lambda_j^\prime/\lambda_j = x$ in the integrals of Eq.~\ref{eq:moment-limit} and after
some tedious calculations, we are able to show that
\begin{equation}\label{eq:part2}
    \sum_{j=1}^{2}\int_{-\pi}^{\pi}\frac{dk}{2\pi}
    \Big( i\frac{\lambda_j^{\prime}}{\lambda_j}\Big)^r
    \Big\vert \langle \lambda_j \vert \tilde{\Psi}(0, k) \rangle  \Big\vert^2
    =
    \int_{-\sqrt{\frac{\tau}{\tau+2}}}^{\sqrt{\frac{\tau}{\tau+2}}} x^rf(x) dx,
\end{equation}
where $f(x)$ satisfies
\begin{widetext}
\begin{equation}\label{eq:f-x}
f(x) =  \frac{1}{\pi(1 - x^2)\sqrt{2\tau - 2(\tau+2)x^2}}
        \Big\{   1 + 2\mathfrak{R}(\alpha^\dag\beta)
              + 2(\vert\beta\vert^2-\vert\alpha\vert^2)x
              + \Big(1 - 2\mathfrak{R}(\alpha^\dag\beta)\frac{\tau+4}{\tau}\Big)x^2 \Big\}.
\end{equation}
\end{widetext}
Substitute Eq.~\ref{eq:part1} and \ref{eq:part2} into Eq.~\ref{eq:moment-limit},
we obtain the $r$-th moment of the quantum walker’s probability distribution
\begin{eqnarray*}
  \lim_{t \rightarrow \infty} \mathbb{E}\Big[\Big(\frac{X_t}{t}\Big)^r \Big] &=&
        \hat{P} + \int_{-\sqrt{\frac{\tau}{\tau+2}}}^{\sqrt{\frac{\tau}{\tau+2}}} x^rf(x) dx \nonumber\\
  ~ &=& \int_{-\infty}^{\infty}
        x^r\Big\{ \delta_{0,x}\hat{P} + f(x)I_{(-\Omega, \Omega)}(x) \Big\}dx.
\end{eqnarray*}
\end{proof}

\begin{figure}
  \centering
  \includegraphics[width=0.45\textwidth]{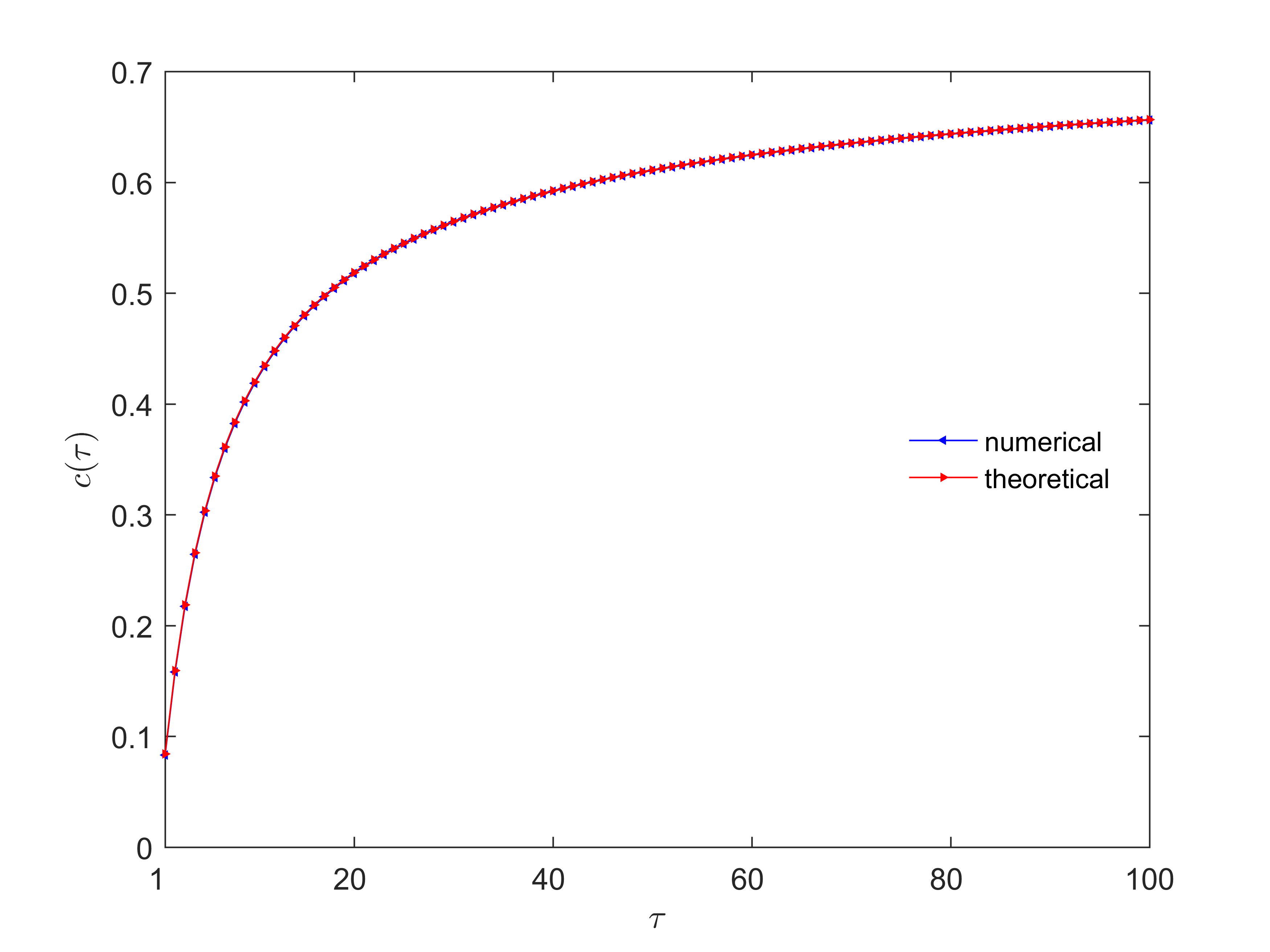}
  \caption{Numerical and theoretical spread coefficients for different laziness factors.
            The system initial state is $\alpha=\frac{1 + \sqrt{2}i}{2}$, $\beta=\frac{\sqrt{2}(1 + i)}{4}$.
            This state is designated to guarantee that
            both $\mathfrak{R}(\alpha^\dag\beta)$ and $\vert\beta\vert^2-\vert\alpha\vert^2$ do not equal to $0$.
            The theoretical results are obtained by Eq.~\ref{eq:coe-solution}.
            The numerical results are got by fitting the numerical data to function $ct^\alpha$ in Matlab.
            The slight difference between two curves is due to that
            the numerical data is obtained by running the walks for only 1000 steps.}\label{fig:coe}
\end{figure}

As a corollary of the weak limit theorem, we prove all LQWs
with system initial states defined in Eq.~\ref{eq:init-state} spread ballistically
and obtain an analytical expression for the variance of the walker's probability distribution.
The variance of a walker's probability distribution is defined as
$$
\sigma^2 = E(X_t^2) - E^2(X_t) \sim ct^\alpha,
$$
where $c$ is the spread coefficient, and $\alpha$ is the spread exponent.
The spread behavior of a quantum walk is determined by the spread exponent of the variance.
If $\alpha = 2$, we say that the walk spreads ballistically;
If $\alpha = 1$, we say that the walk spreads diffusively.
It has been proved that for standard quantum walks $\alpha = 2$~\cite{chandrashekar2008optimizing},
while for random walks $\alpha = 1$~\cite{venegas2012quantum}.
\begin{corollary}\label{thm:coe}
  For a LQW whose laziness factor is $\tau$,
  if the walker starts with the system initial state given in Eq.~\ref{eq:init-state},
  the variance of the walker's probability distribution satisfies
  \begin{equation*}
    \sigma^2 = c(\tau, \alpha, \beta) t^2,
  \end{equation*}
  where $c(\tau, \alpha, \beta)$ is the spread coefficient defined in Eq.~\ref{eq:coe-solution}.
\end{corollary}
We can see easily from Corollary~\ref{thm:coe} that all lackadaisical quantum walks spread ballistically
for system initial states defined in Eq.~\ref{eq:init-state} as the spread exponent is $2$.
Moreover, we obtain an analytical solution for the spread coefficient $c(\tau, \alpha, \beta)$ of the variance
in Eq.~\ref{eq:coe-solution}, from which we find that it is dependent on both $\tau$ and coin initial state $\alpha$, $\beta$
and the laziness factor $\tau$.
By tuning the parameters $\tau$, $\alpha$ and $\beta$, we can achieve arbitrary spread coefficients
in the range $(0, 1)$.
We conduct numerical simulations to calculate the spread coefficients for different laziness factors and
the comparison between numerical and theoretical results are illustrated in Fig.~\ref{fig:coe}.
\begin{proof}
By Theorem~\ref{thm:weak-limit} it is easy to see that
\begin{eqnarray*}
  \sigma^2 &=& E(X_t^2) - E^2(X_t) \\
  ~ &\sim& t^2\int_{-\Omega}^{\Omega} x^2f(x) dx - t^2 \Big\{\int_{-\Omega}^{\Omega} xf(x) dx\Big\}^2 \\
  ~ &=& c(\tau, \alpha, \beta) \; t^2,
\end{eqnarray*}
where the coefficient function $c(\tau, \alpha, \beta)$ is defined as
\begin{equation}\label{eq:coe-expression}
  c(\tau, \alpha, \beta) = \int_{-\Omega}^{\Omega} x^2f(x) dx - \Big\{\int_{-\Omega}^{\Omega} xf(x) dx\Big\}^2.
\end{equation}
Solving Eq.~\ref{eq:coe-expression}, we obtain the analytical solution for $c(\tau, \alpha, \beta)$ which has the form of
\begin{widetext}
\begin{equation}\label{eq:coe-solution}
c(\tau, \alpha, \beta) =
    1 - \frac{(5\tau+8)\sqrt{2\tau+4}}{(2\tau+4)^2}
  + \Bigg\{  \frac{2(\tau^2+12\tau+16)\sqrt{2\tau+4}}{\tau(2\tau+4)^2} - \frac{4}{\tau}  \Bigg\}\mathfrak{R}(\alpha^\dag\beta)
  - \Bigg\{\Big(1 - \frac{\sqrt{2\tau+4}}{\tau+2}\Big)(\vert\beta\vert^2-\vert\alpha\vert^2) \Bigg\}^2.
\end{equation}
\end{widetext}
\end{proof}

\section{Conclusion}\label{sec:conclusion}

In this paper, we analyze in detail the properties of LQWs on a line
for arbitrary laziness factor $\tau$.
First, we study the localization phenomenon shown in the walks.
With the discrete Fourier transformation method,
we are able to present an explicit form for the localization probability of the walker 
in the limit of $t \rightarrow \infty$.
The limiting coin state is obtained by a set of linear maps $F_j$ on the initial coin state.
This set of $F_j$ contain all information required to depict the walker's behavior at the origin.
The localization probability is the inner product of the limiting coin state,
which is shown independent on the initial coin state, and totally determined by $\tau$.
We also calculate the velocities of the left and right-travelling probability peaks
appeared in the walker's probability distribution.
We can control the spread behavior the quantum walks and achieve faster spreading than
the standard quantum walks by tuning the laziness factor.
Furthermore, we show that when $\tau$ approaches infinity, the LQW degenerates to a trivial walk.
At last, we calculate rigorously the system state
and get a long time approximation for the entire probability density function.
The density function has both the Dirac $\delta$-function and a continuous function
with a compact support whose domain is determined by the peak velocities.
As an application of the density function, 
we prove that all LQWs spread ballistically, 
and give an analytic solution for the variance of the walker's probability distribution.
The analytical results we obtain illustrate interesting characteristics of
LQWs compared to standard quantum walks and the corresponding lazy random walks.
For example, it is obvious that the greater the $\tau$ is, the more the walker prefers to stay in lazy random walks.
However, a lackadaisical quantum walker spread even faster with the increment of $\tau$.
That's why we say the lackadaisical quantum walker is \textbf{not} lazy at all.

\section*{Acknowledgement}

The authors want to thank Haixing Hu, Qunyong Zhang, Xiaohui Tian and Huaying Liu for
the insightful discussions. K. W. wants to thank Takuya Machida for his kind help.
This work is supported by the National Natural Science Foundation of China
(Grant Nos. 61300050, 91321312, 61321491) and
the Chinese National Natural Science Foundation of Innovation Team (Grant No. 61321491).



\end{document}